\documentclass[manuscript]{aastex}

\begin{document}

\title{Thermal Evolution of Strange Stars}
\author{Zhou Xia}
\affil{The Institute of Astrophysics, HuaZhong Normal University,
Wuhan 430079, China.} \email{zhoux@phy.ccnu.edu.cn}
\author{Wang Lingzhi}
\affil{College of Physical and Technology, HuaZhong Normal
University, Wuhan 430079, China.}
 \and
\author{Zhou Aizhi}
\affil{College of Physical and Technology, HuaZhong Normal
University, Wuhan 430079, China.}

\begin{abstract}
We investigated the thermal evolution of rotating strange stars with
the deconfinement heating due to magnetic braking. We consider the
stars consisting of either normal quark matter or
color-flavor-locked phase. Combining deconfinement heating with
magnetic field decay, we find that the thermal evolution curves are
identical to pulsar data.
\end{abstract}

\section{Introduction}

Combining observational data available from the successful launch of
Chandra and XMM/Newton X-ray space missions with the significant
progress in theoretical studies of the physics of dense matter,
these developments offer the hope for distinguishing various
competing neutron star thermal evolution models. Through this way we
would have an opportunity to determine various important properties
of dense matter, such as the composition, superfluidity and the
equation of state. These progresses should help us obtain deeper
insight into the properties of dense matter.

Quantum chromodynamics(QCD) predicted the existence of quark matter
at high density. Moreover phenomenological and microscopic studies
also confirmed that quark matter at a sufficiently high density, as
in compact stars, undergoes a phase transition into a color
superconductivity state, which are typical cases of the 2-flavor
color superconductivity (2SC) and color-flavor locked (CFL) phases.
Theoretical approaches concur that the superconductivity order
parameter, which determines the gap in the quark spectrum, lies
between 1 and $100Mev$ for baryon densities existing in the
interiors of compact stars. It is generally believed that appearance
of quark matter would be implied in structure and evolution of
compact stars. Mass-radius relation and changes in structure have
been deeply investigated in an amount of
literatures\citep{zheng06b,alcock86,Glendenning95a,
Glendenning95b,yang02,pan07}. The evolutionary properties associated
with spin of the star have been extensively discussed in past
work\citep{bildsten00,madsen00,andersson00,pan06a,pan06b,liu04,kang05,zheng07}.
Meanwhile, someones had ever expected compact star cooling as a
probe to distinguish quark matter from hadronic
matter\citep{liu05,zheng06a}. The researches showed that the cooler
stars contain perhaps quark matter. However, the problem is
unsolvable. As known, continuous deconfinement processes occur
during the rotation evolution of the star if quark core
exists\citep{Zdunik01,alford04,kang07}. The corresponding
deconfinement heating(DH) extremely influences the evolution
temperature of the star. No matter whether a strange star which
sustain a tiny nuclear crust is in normal phase or in color
superconducing phase, DH should delay the cooling of the star.
Consequently, the strange star become warmer so that it wouldn't
cooler\citep{yu04,zhou06}.

It is well know that a compact star spins down due to magnetic
dipole radiation. So the DH processes are closely related with the
magnetic field. When we take constant field into account for the
calculations of heating rate, the star would maintain too high
temperature at old age evolutionary time, during which we have not
found such hot pulsar yet(see figures in \citep{yu04,zhou06}) except
a few of millisecond pulsars. However, the observations support the
fact that the field could decay such as Ohmic decay and the decay
time scale is about $10^6$ years. Obviously, the DH will decrease
due to field decay at old ages. In this paper, we will assess how
the thermal evolution of a strange star would be affected.

\section{Neutrino Emissivities and Specific Heat}

The most efficient cooling process in unpaired quark matter is the
quark direct Urac(QDU) process $d\rightarrow ue\bar\nu $ and
$ue\rightarrow d\bar\nu $, given by \citep{Iwamoto82}
\begin{equation}
\epsilon^{(D)}\simeq8.8\times10^{26}\alpha_{c}uY^{1/3}_{e}T^{6}_{9}\zeta_{D}
~ erg cm^{-3}sec^{-1}.
\end{equation}
where $\alpha_{c}$ is the strong coupling constant,
$u=\rho_{b}/\rho_{0}$,$\rho_{b}$ is the baryon density and $\rho_{0}
= 0.17 fm^{-3}$ is the nuclear saturation
density,$Y_{e}=\rho_{e}/\rho_{b}$ is the electron fraction, and
$T_{9}$ is the temperature in units of $10^{9}$K. When the QDU
process being switched off due to a small electron fraction
($Y_e<Y_{ec}=(3/\pi)^{1/2}m_e^3\alpha_c^{-3/2}/64$.), the dominating
contribution to the emissivities is the quark modified Urca(QMU)
$dq\rightarrow uqe\bar\nu$ and quark bremsstrahlung(QB)
$q_1q_2\rightarrow q_1q_2\nu\bar\nu $ processes, estimated
as\citep{Iwamoto82}
\begin{equation}
\epsilon^{(M)}\simeq2.83\times10^{19}\alpha_{c}^{2}uT^{8}_{9}\zeta_{M}
~ erg cm^{-3}sec^{-1},
\end{equation}
\begin{equation}
\epsilon^{(QB)}\simeq2.98\times10^{19}uT^{8}_{9}\zeta_{QB} ~ erg
cm^{-3}sec^{-1}.
\end{equation}

Because of the pairing in color superconducting phase, the
emissivity of QDU process is suppressed by a factor of $\zeta_D \sim
exp(-\Delta/T)$ and the emissivities of QMU and QB processes are
suppressed by a factor $\zeta_M \sim exp(-2\Delta/T)$ for
$T<T_{c}$\citep{Blaschke00,Shovkovy04}.
\par
In order to compute the cooling curves of the stars,we need to give
the specific heat of the electrons and quarks\citep{Iwamoto82}:
\begin{equation}
c_{e}\simeq 2.5\times10^{20}u^{2/3}T_{9} ~ erg cm^{-3}K^{-1}
\end{equation}
\begin{equation}
c_{q}\simeq0.6\times10^{20}Y^{2/3}_{e}u^{2/3}T_{9} ~ erg
cm^{-3}K^{-1}
\end{equation}
\par
But in color superconductivity phase, the quark specific heat  is
changed exponentially\citep{Blaschke00}
\begin{equation}
c_{sq}=3.2c_{q}\left(\frac{T_{c}}{T}\right)\times\left[2.5-1.7\left(\frac{T}{T_{c}}\right)+3.6\left(\frac{T}{T_{c}}\right)^{2}\right]\exp\left(-\frac{\Delta}{k_{B}T}\right)
\end{equation}
where $T_{c}$ is critical temperature related to $\Delta$ as
$\Delta=1.76T_{c}$. Compare with the total mass of the stars, the
mass of the crust is very small($M_{c}\leq 10^{-5}M_{\odot}$). So we
neglect its contribution to neutrino emissivity and specific
heat\citep{Lattimer94}.

\section{DH with Magnetic-field decay}

We consider a strange star embodied by a nuclear crust. The DH is
determined by the mass change of the crust. The total heat released
per unit time as a function of $t$ is:
\begin{equation}
H_{\rm dec}(t)=-q_n\frac{1}{m_{b}}\frac{d M_{\rm c}}{d
{\nu}}\dot{\nu},
\end{equation}
where $q_n$, the heat release per absorbed neutron, is expected to
be in the range $q_n{\sim}10-40{\rm MeV}$. Its specific value
depending on the assumed SQM model, and $m_{b}$ is the mass of
baryon. The mass of the crust $M_{\rm c}$ can be approximated by a
quadratic function of rotation frequency $\nu$. As discussed in
\citep{Zdunik01,yu04}, the mass of the crust reads
\begin{equation}
M_{\rm c}=M^{0}_{\rm c}(1+0.24\nu^{2}_{3}+0.16\nu^{8}_{3})
\end{equation}
where $\nu_{3}=\nu/10^{3}$Hz and $M^{0}_{\rm c}\leq 10^{-5}M_\odot$
is the mass of the crust in the static case.

Assuming the spin-down is induced by the magnetic dipole radiation,
the evolution of the rotation frequency $\nu$ is given
by\citep{yu04}
\begin{equation}\label{nu}
\dot{\nu}=-\frac{2\pi^{2}}{3Ic^3}B(t)^2R^6\nu^3{\rm sin}^2{\theta},
\end{equation}
where $I$ is the stellar moment of inertia, $\theta$ is the
inclination angle between magnetic and rotational axes. In our work,
we combined the heating with magnetic field decay, so in
Eq.(\ref{nu}) the magnetic is denoted as a function of time.

The field decay of the magnetic field is expected to be a very
complicated process. For purposes of illustrate its main features,
we assume a simple model equation\citep{Jose07},
\begin{equation}\label{bt}
\frac{dB}{dt}=-\frac{B}{\tau_{D}}
\end{equation}
with the initial condition $B(0)=B_0$. Eq.(\ref{bt})is a crude
approximation, we are dealing with a typical value of the magnetic
field, neglecting the spatial variation of the field throughout the
crust. Moreover, we adopt that the magnetic field dissipate on a
time scale $\tau_D$
\section{Cooling Curves}

Considering the energy run away and heating effect of the star, the
cooling equation can be written as:
\begin{equation}
C_{\rm V}\frac{d T}{d t}=-L_{\nu}-L_{\gamma}+H,
\end{equation}
where $C_{\rm V}$ is the total specific heat, $L_{\nu}$ is the total
neutrino luminosity and $L_{\gamma}$ is the surface photon
luminosity given by
\begin{equation}\label{ls}
L_{\gamma}=4{\pi}R^2{\sigma}T_s^4,
\end{equation}
where ${\sigma}$ is the Stefan-Boltzmann constant and $T_s$ is the
surface temperature. The last term in Eq.(\ref{ls}) represents the
DH due to the spin-down of the star.

The surface temperature of the stars is related to internal
temperature by a coefficient determined by the scattering processes
occurring in the crust. We apply an formula which is demonstrated by
\citep{Gudmundsson83}. It reads
\begin{equation}\label{ts}
T_{s}=3.08\times10^6g_{s,14}^{1/4}T_{9}^{0.5495},
\end{equation}
where $g_{s,14}$ is the proper surface gravity of the star in units
of $10^{14}\rm cm\hspace{0.1cm}s^{-2}$\citep{Potekhin01}. In
principle, magnetic fields may change the expression of
Eq(\ref{ts}). However, \citep{Potekhin01} have present that the
effect is negligible if the field strength is lower than $10^{13}G$.
So Eq(\ref{ts})is a good approximation for our case.

Considering the gravitational red-shift, and then the effective
surface temperature detected by a distant observer is
\begin{equation}
T_{s}^{\infty}=T_{s}\left[1-0.295\left(\frac{M}{M_{\odot}}\right)R_6^{-1}\right]^{1/2}
\end{equation}
here $R_6$ is the radium of the star in units of $10^6~cm$

We consider a canonical strange star of $1.4M_{\odot}$ at a constant
density in our work, which is a very good approximation for strange
stars of mass $M\leq1.4M_{\odot}$\citep{alcock86}. We choose $u=3$,
$q_n=20 ~ Mev$, the initial temperature $T_0=10^{9} ~ K$, initial
period $P_0=0.78 ~ ms$, initial mass of the crust
$M_c^0=10^{-5}M_{\odot}$, the magnetic tilt angle
$\theta=45^{\circ}$ and the time scale $\tau_D=2\times 10^6 ~ yr$.

Using the model described in the preceding section, we plot the
cooling curves of rotating strange star in normal phase with DH for
various initial magnetic fields ($10^{11}-10^{13} ~ G$) in Fig.1.
The observational data are taken from \citep{Page04}. We take
$Y_e=10^{-5}$ for $Y_e>Y_{ec}$ which is a representative for the QDU
process contributing to the cooling, whereas $Y_e=0$ for
$Y_e<Y_{ec}$ when QMU and QB processes dominates.

We also show the thermal evolution curves of strange star in CFL
phase in Fig.2. In contrast to previous work\citep{Blaschke00}, DH
increases surface temperature of strange stars effectively. With
constant fields, the stars still maintain high temperature at old
ages but we can't find such hot pulsars during the phase. In the
case of field decay, the cooling curves are exponentially suppressed
at tails. This modification is very important. Under consideration,
the thermal evolutionary model is compatible with the pulsar data.
\section{Conclusion}
We have given out the cooling curves of rotating strange stars by
considering the DH effect with magnetic field decay. The thermal
evolution of rotating strange stars are different from previous
works. The DH could increase the temperature of strange star to the
inferred pulsar data while the magnetic field decay suppresses DH to
retain rapid cooling at old ages($t>10^6~yr$). Obviously, there is
no evidence for the existence of extra hot source at old age.

We should noticed that the used model of magnetic field decay which
is crude. We here apply a simple magnetic field decay model. The
decay time scale is taken according to statistical data of pulsar.
Actually, the magnetic field decay model is an endless controversial
problem. The rigorous discusses may be necessary but our conclusion
won't be changed by future study.
\acknowledgments

This work was supported by the NFSC under Grant No. 10603002.

\clearpage

\begin{figure}
\includegraphics{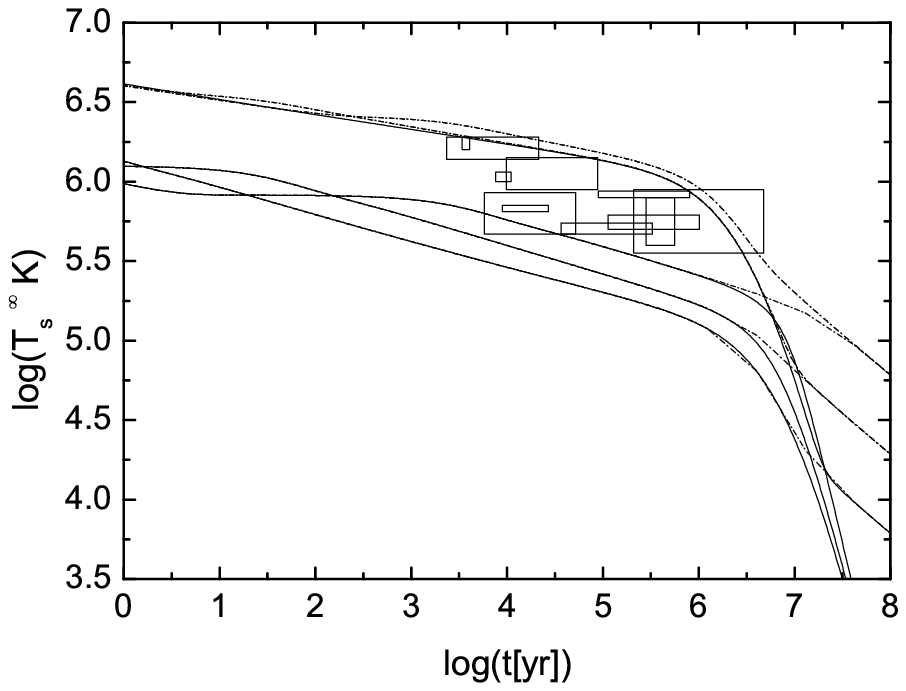} \caption{Cooling curves of rotating strange star in
normal phase. The up group of lines are in the case which only mDURA
process occured, the lower group of lines are in the case which DURA
process switched on. The solid lines correspond to magnetic field
decay effect. The dash-dotted lines represent the constant field
case.}\label{fig1}
\end{figure}

\begin{figure}
\includegraphics{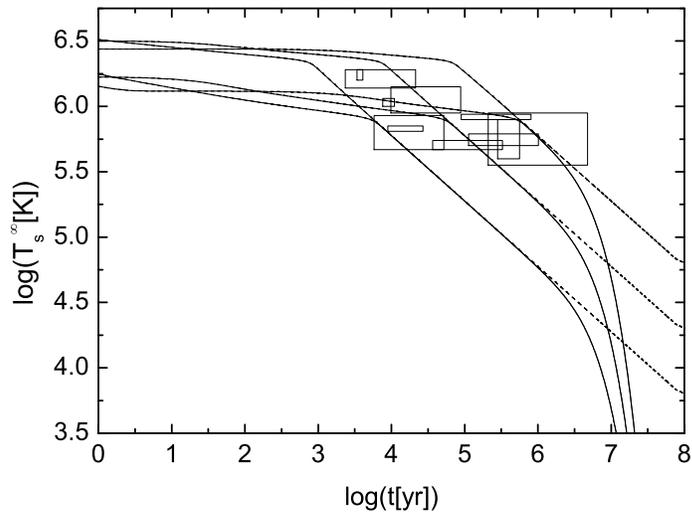}\caption{Cooling curves of rotating strange star in
CFL phase. The up group of lines are in $\Delta=1 Mev$, the lower
group of lines are in $\Delta=0.1 Mev$. The solid lines correspond
to magnetic field decay effect. The dotted lines represent constant
field case.}\label{fig2}
\end{figure}

\end{document}